\documentclass[conference]{IEEEtran}
\usepackage{amsmath,amsfonts}
\usepackage{algorithmic}
\usepackage{array}
\usepackage[caption=false,font=normalsize,labelfont=sf,textfont=sf]{subfig}
\usepackage{textcomp}
\usepackage{stfloats}
\usepackage{url}
\usepackage{verbatim}
\usepackage{graphicx}
\usepackage[colorlinks,linkcolor=blue,anchorcolor=blue,citecolor=blue,urlcolor=blue,bookmarks=false]{hyperref}
\usepackage{balance}
\usepackage{cite}
\usepackage{orcidlink}

\usepackage{tikz,pgfplots,pgf}

\usepackage{bbm}
\usepackage{soul,xcolor}

\usepackage{amssymb}
\usepackage{amsmath}
\usepackage{amsthm}
\usepackage[shortlabels]{enumitem}
\usepackage{mathtools}
\usepackage{dcolumn}
\usepackage{physics}
\usepackage{float}
\usepackage{bm}
\usepackage{braket}
\usepackage[normalem]{ulem}
\usepackage[ruled,vlined,linesnumbered]{algorithm2e}
\usepackage{thm-restate}

\usepackage{stackengine}
\usepackage{qcircuit}

\usepackage{diagbox}

\usepackage{cuted}

\usepackage{fancyhdr}

\usepackage{xr-hyper}
\makeatletter

\pgfplotsset{compat=1.16}
\usepackage[T1]{fontenc} % Ensures font encoding
\IEEEoverridecommandlockouts

\begin{document}

\title{Implementing Finite Impulse Response Filters on Quantum Computers
\thanks{The project is funded by the U.S. Department of Energy, Advanced Scientific Computing Research, under contract number DE-SC0025384.}
}

\author{{Aishwarya Majumdar\orcidlink{0009-0008-2800-0455} \quad Bojko N. Bakalov\orcidlink{0000-0003-4630-6120} \quad Dror Baron\orcidlink{0000-0002-6371-8496} \quad Yuan Liu\orcidlink{0000-0003-1468-942X} }

\IEEEauthorblockA{%\textit{Electrical and Computer Engineering} \\
\hfill \\
\textit{North Carolina State University,}
\textit{Raleigh, North Carolina 27695, United States} \\
\{amjumd4, bnbakalo, barondror,
q\_yuanliu\}@ncsu.edu
}
\vspace{-1\baselineskip}
}

\maketitle

\vspace{-1\baselineskip}

\begin{abstract}
While signal processing is a mature area, its connections with quantum computing have received less attention.
In this work, we propose approaches that perform classical discrete-time signal processing using quantum systems. Our approaches encode the classical discrete-time input signal into quantum states, and design unitaries to realize classical concepts of finite impulse response (FIR) filters. We also develop strategies to cascade lower-order filters to realize higher-order filters through designing appropriate unitary operators.
Finally, a few directions for processing quantum states on classical systems after converting them to classical signals are suggested for future work.
\end{abstract}
\vspace{-0.5\baselineskip}

%\vspace{-1\baselineskip}

\section{Introduction}
\IEEEPARstart{S}{ignal} Processing emerged as a field of utmost importance more than 75 years ago \cite{SPHistory}. Since then, the rapid development of modern signal processing techniques for both digital \cite{10.5555/227373,oppenheim1999discrete} and analog \cite{10.5555/248702} signals has produced many important areas that are intertwined with almost all fields of modern technologies, including computing~\cite{cormen2022introduction}, communications \cite{heath2017introduction}, control~\cite{bishop2011modern}, 
%sensing \cite{}, 
and machine learning \cite{intersectionsignalprocessingmachine}. 

While quantum computing emerged as a field of interest much later \cite{feynman1985quantum}, it has seen significant growth in the past decades with advancements in quantum theory \cite{knill1997theory,roffe2019quantum,vandersypen2004nmr}, algorithms \cite{bharti2022noisy,grandunification}, and hardware development \cite{bruzewicz2019trapped,bravyi2022future,henriet2020quantum,ColorCenters10.1063/5.0007444,RevModPhys.95.025003}. Among these developments, the generalization of the concept of signal processing from classical to quantum systems \cite{eldar2002quantum} has produced an important class of quantum algorithms, i.e., quantum signal processing (QSP) \cite{low2016methodology,low2017optimal}, and its high-dimensional generalization, the quantum singular value transform \cite{Gilyen2019}. The concept of filter design in classical signal processing has since been used in designing quantum algorithms that can realize polynomial transformations of quantum amplitudes, leading to a grand unification of quantum algorithms with advantages in many applications \cite{grandunification}. More generalizations of QSP 
to realize a wider class \cite{motlagh2023generalized,dong2022infinite} or a larger number of polynomials \cite{lu2024quantum,laneve2023quantum}, generalization from single to multiple variables \cite{rossi2021mqsp,nemeth2023variants,laneve2024multivariate}, from digital to analog \cite{rossi2023quantum}, and mixed analog-digital quantum systems \cite{qspi2023,liu2024toward}, have recently been proposed, tantalizing researchers with even broader applications of these quantum algorithms. 

Despite these parallel developments in both classical and quantum domains of signal processing algorithms, perhaps surprisingly, the connection between these two fields has been largely missing. As we envision a future of classical and quantum computers co-existing for joint signal processing of classical and quantum signals, two important questions immediately arise. Given the advantage of quantum computers for many problems \cite{Nielsen_Chuang}, are there advantages to performing classical digital signal processing on quantum computers? On the flip side, are there advantages to processing quantum signals on classical computers?
Addressing these two questions is crucial to understanding the fundamental limits and advantages of both classical and quantum signal processing techniques, in order to best combine them for practical applications. 

Despite the simplicity of these questions, answering them is highly non-trivial. For example, one of the pillars of signal processing is the Fourier transform. The computational advantage of the quantum Fourier transform \cite{moore2006generic} over the classical fast Fourier transform \cite{van1992computational} seems to suggest that performing classical signal processing on quantum computers is advantageous. However, encoding classical time-domain signals as quantum states may be costly. Moreover, 
when filtering time series signals, classical signal processing arithmetic operations are often non-unitary, and it is not clear how to design them using the unitary operations required by quantum computers.
Conversely, for classical processing of quantum signals, it is unknown how to do the quantum to classical signal conversion, as well as what classical signal processing filters should be used to achieve desired quantum processing tasks. We note that a recent work \cite{sequencyHadamart} proposed a quantum algorithm for high- and low-pass filtering using a sequency-ordered Walsh-Hadamard transform for a given input state. However, it is not clear how to realize a general filter nor how to encode classical signals as quantum states. To answer these questions, a unified framework that combines classical and quantum signal processing is required. 

\begin{figure}
    \centering
\includegraphics[width=0.35\textwidth]{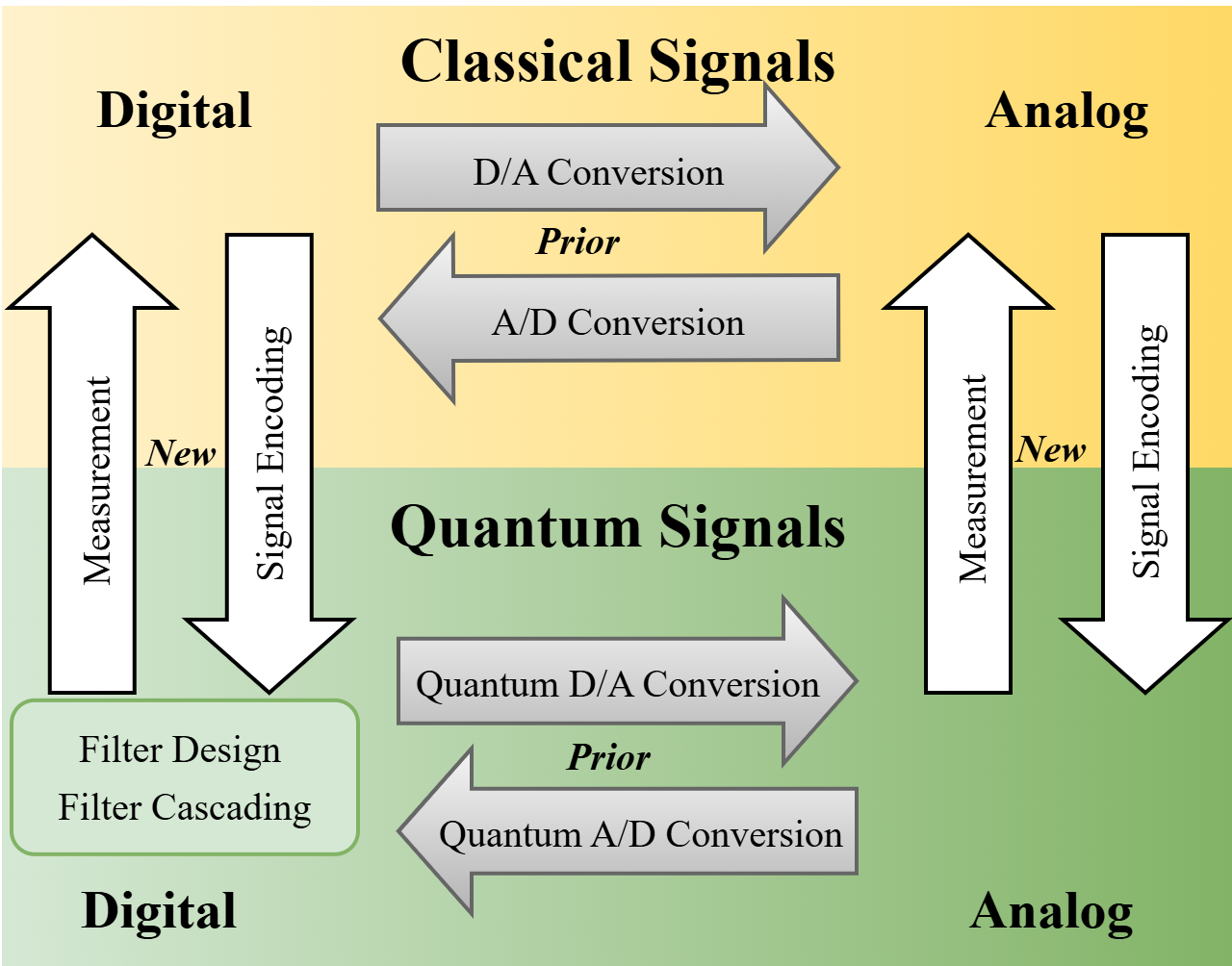}
    \caption{Interconnections between various signal processing domains.}
    \label{fig:signal-processing-overview}
    \vspace*{-1.5\baselineskip}
\end{figure}

In this work, we address these questions by establishing connections between classical and quantum signal processing. Fig.\ref{fig:signal-processing-overview} provides a high-level schematic overview of our approach to unifying classical and quantum signal processing. 
The upper part of Fig.\ref{fig:signal-processing-overview} shows the prior art of classical signal processing techniques, which connects digital and analog signals. The lower part illustrates recent parallel developments in the quantum domain for mixed analog-digital QSP \cite{liu2024toward,qspi2023,liu2024hybrid}. The main contributions of our work are indicated by the vertical arrows, which establish a connection between classical and quantum signal processing. 

The rest of this paper is organized as follows. Sec. \ref{sec:classical2quantum} discusses strategies for designing appropriate unitaries and operators that realize on a quantum computer classical filters and filter-cascading techniques commonly used in classical signal processing. 
We present results from running examples of these quantum circuits on IBM Qiskit's QASM simulator.
For processing quantum signals with classical signal processing,
in order to lay out a framework for future investigations,
we describe open problems on the potential advantages 
in Sec. \ref{sec:quantum2classical}.
We conclude the paper with short discussions in Sec. \ref{sec:dis-conclusion}. 

%%%%%%%%%%%%%%%%%%%%%%%%%%%%%%%%%%%%
\section{Classical to Quantum}
\label{sec:classical2quantum}
 
We discuss the processing of classical signals on quantum computers in this section. In Sec. \ref{ssec:classical-quantum-encoding}, we first establish a quantum encoding of any classical signals. Sec. \ref{ssec:quantum-filter} then presents strategies to perform time-domain signal processing for filter design and cascading on quantum computers. 

\subsection{Classical Digital to Quantum Digital Converter}
\label{ssec:classical-quantum-encoding}

In the quantum paradigm, the fundamental processing unit is a qubit, which can be initialized to a superposition (i.e., a complex linear combination) of two states, $\ket{0} = [1,0]^T$
and $\ket{1} = [0,1]^T$, given that the modulus squared of the coefficients sum to 1. Having several qubits corresponds to taking a tensor product of vectors, and $m$ qubits are encoded as a vector in $(\mathbb{C}^2)^{\otimes m} = \mathbb{C}^{2^m}$ with a unit norm.
In closed quantum systems, applying any valid transformation to these qubit states is accomplished through unitary operations, which are norm-preserving.

Consider a discrete-time signal to be $\{x[0], x[1], \dots, x[n-1], x[n], x[n+1], \dots \}$, such that the $n$-th time sample is given by $x[n]$. To encode $x[n]$ as an amplitude in a quantum state, we first uniformly normalize the signal and denote such samples as $\{x_0, x_1, \dots, x_{n-1}, x_n, x_{n+1}, \dots \}$.  
One approach to normalize the samples could be to assume that the maximum amplitude of the samples is upper bounded by known $M$, and 
the system simultaneously processes $d$ samples. Then every sample is scaled by a factor of $\frac{1}{M\sqrt{d}}$, so that each of the basis state coefficients has amplitude less than 1 and the signal shape is preserved.
If the quantum signal processing system processes only one sample at a time, the  discrete-time signal can be encoded to a qubit state as $\ket{X_n} = \begin{bmatrix}
    x_{n} \\
    \sqrt{1-|x_{n}|^2}
    \end{bmatrix}$,
which can be prepared by applying rotation gates to the qubit. In systems that simultaneously process $d$ samples at a time, the input quantum state requires
$m\geq \lceil \log_2{(d+1)} \rceil$ qubits to encode the samples,
and the  quantum input state would be 
\begin{align}\label{input-state-xn}
% \scalebox{0.9}{$
    \ket{X_n} = \left[
        x_{n-(d-1)}, 
        \cdots, 
        x_{n},   
        0,       
        \cdots,
        0,
        \eta_n
    \right]^T, 
    % $}
\end{align}
where $\eta_n = \sqrt{1-(\sum_{i=0}^{d-1}|x_{n-i}|^2)}$ is a normalization term. 
We add zeros in \eqref{input-state-xn} to make the dimension of the vector $2^m\ge d+1$.
For filters, $d$ denotes the number of filter taps. 
While the scaling is logarithmic, the value of $d$ is limited by the current hardware capabilities, requiring coherent and non-dissipative quantum gates.

%%%%%****************************%%%%%%%

\subsection{Processing Quantum Signals in Time Domain}
\label{ssec:quantum-filter}

\subsubsection{Filter Design, but Quantumly}
% Now we look at designing unitary operators to process the quantum signal states on quantum computers. 
To design unitary operators for processing classical signals on quantum computers, we consider a simple case of a linear-phase finite impulse response (FIR) filter represented by the difference equation 
\begin{align}\label{filter-CD-difference-eqn} 
% \scalebox{0.9}{$
    y[n] = \sum_{i=0}^{d-1}p_{i}\,x[n-i]. 
    % $}
\end{align}
To perform signal processing quantumly, the following steps are adopted.

\textit{Input State Initialization}: 
    Since the filter considered processes $d$ samples, the
    input state is as defined in \eqref{input-state-xn}.
\par
\textit{Unitary Design}: The filter unitary $U$ is designed such that the first $d$ elements in the $d$-th row are normalized coefficients of the filter such as
    \begin{align} \nonumber
    % \scalebox{0.9}{$
        U = \begin{bmatrix}
            * & \cdots & * & \cdots & * & 0\\
            \vdots & \vdots & \vdots & \vdots & \vdots & \vdots \\
             \frac{p_{d-1}}{\sqrt{\sum_{i=0}^{d-1}p_{i}^2}} & \cdots & \frac{p_0}{\sqrt{\sum_{i=0}^{d-1}p_{i}^2}} & 0 &\cdots & 0 \\
             * & \cdots & * & \cdots & * & 0\\
             \vdots & \vdots & \vdots & \vdots & \vdots & \vdots \\
             0 & \cdots & 0 & 0 & \cdots & 1
        \end{bmatrix}.
        % $}
    \end{align}
    The matrix 
    $U$  is of size $2^m \times 2^m$, where $m\ge\lceil \log_2{(d+1)} \rceil$. The blocks denoted by
    $*$ indicate elements that are necessary to make $U$ unitary while not contributing to the computation per se.
    The main challenge in converting a classical filter into a unitary lies in its efficient synthesis using the quantum hardware's native gates. Several unitary decomposition techniques exploit the symmetry and sparsity of the unitaries to generate efficient sequences of single- and two-qubit gates~\cite{Nielsen_Chuang}.
\par 
\textit{Output State}: The output state is computed as 
    \begin{align}\label{qf-input-output-U}
    % \scalebox{0.9}{$
        \ket{Y_n} = U \ket{X_n},
        % $}
    \end{align}
    where $\ket{Y_n}$ is of the form
    \begin{align*}
    % \scalebox{0.9}{$
      \begin{bmatrix}
            *, \cdots,
            \frac{p_{d-1}}{\sqrt{\sum_{i=0}^{d-1}p_{i}^2}}x_{n-(d-1)} + \cdots + \frac{p_0}{\sqrt{\sum_{i=0}^{d-1}p_{i}^2}}x_{n}, 
            *, \cdots
        \end{bmatrix}^T.
        % $}
    \end{align*}
    The filter computation value is obtained at the $d$-th entry of the output state $\ket{Y_n}$.

\textit{Measurement}:
    The projection operator designed to measure $\ket{Y_n}$ can be expressed as $\Pi_{d} = \ket{d} \bra{d}$, where
    $\bra{d} = [ 
         0,
         0,
        \cdots,
         1,
        \cdots,
        0]$ 
        and $\ket{d}$ is its transpose.
    (Note that a single ``1" in $\Pi_{d}$ is positioned at the $d$-th diagonal element.)
On measuring the output state using $\Pi_{d}$, the expected outcome will be $\bra{Y_n}\Pi_{d} \ket{Y_n}$, or $
    \left|\frac{p_{d-1}}{\sqrt{\sum_{i=0}^{d-1}p_{i}^2}}x_{n-(d-1)} + \cdots + \frac{p_0}{\sqrt{\sum_{i=0}^{d-1}p_{i}^2}}x_{n} \right|^2$,
which is a uniformly scaled and squared value of $y[n]$, i.e., the expected filter output corresponding to time sample $n$.
\par 
\textit{Input Update}: Similar to Eq. \eqref{input-state-xn}, the next input state is
\begin{align} \nonumber
\ket{X_{n+1}} = \left[
            x_{n-(d-2)}, 
            \cdots,
            x_{n},
            x_{n+1}, 0, 
            \cdots, 0,
            \eta_{n+1}
        \right]^T.
\end{align}
\vspace*{-1\baselineskip}
\begin{figure}
    \centering
\includegraphics[width=0.45\textwidth]{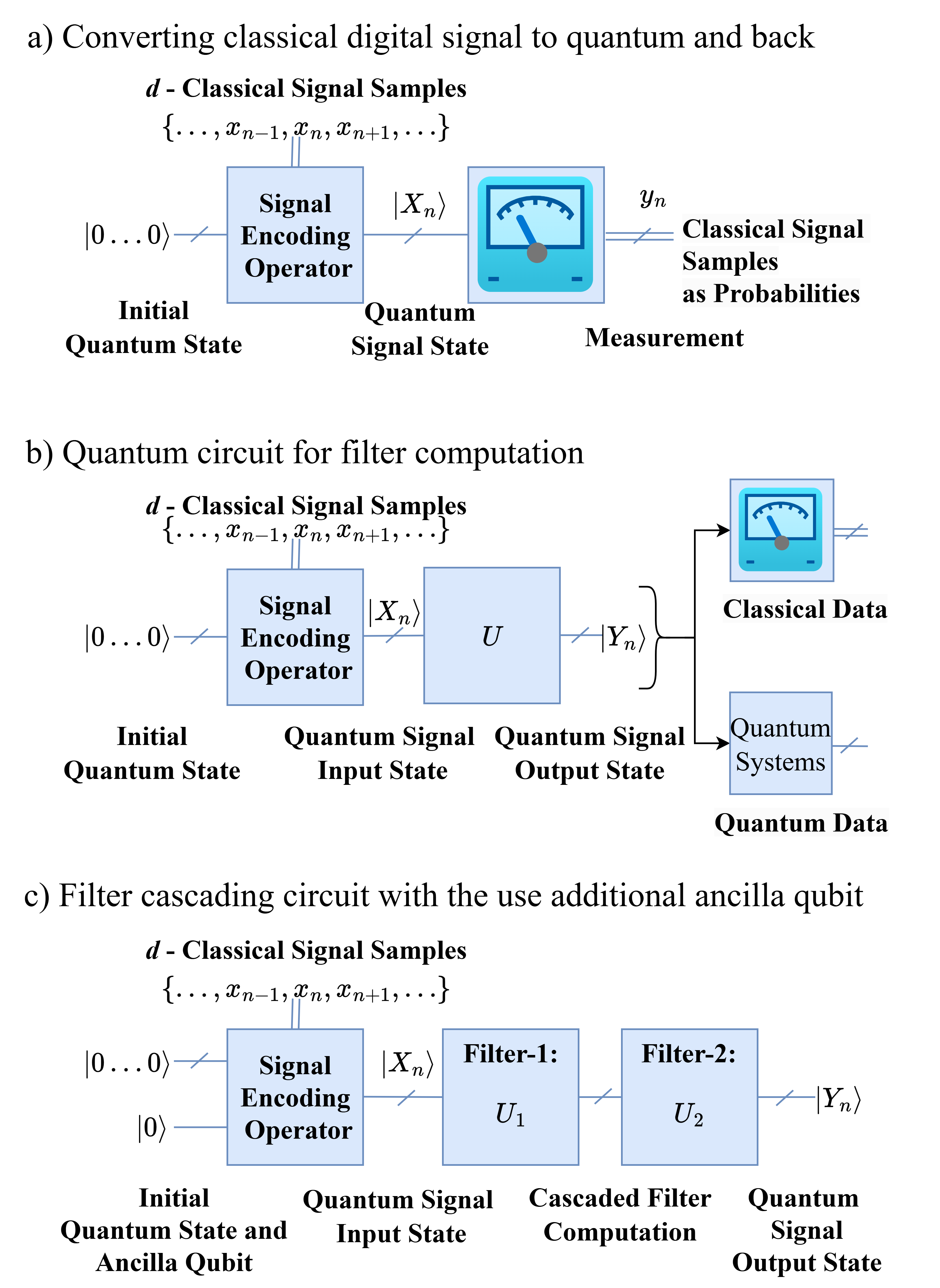}
    \vspace*{-0.5\baselineskip}
    \caption{Various schemes for processing classical digital signals on quantum digital systems. a) classical to quantum signal conversion; b) time-domain processing to implement the filters on quantum computers; c) cascading two filters on quantum computers.}
    \label{fig:quantum-circuits}
    \vspace*{-1.5\baselineskip}
\end{figure}
%%%%%****************************%%%%%%%
\subsubsection{Filter Cascading, but Quantumly}
A common practice in system design is combining lower-order systems to approximate a higher-order system. This section discusses a framework for cascading lower-order filters to approximate a higher-order filter implementable on quantum computers. We consider the following example.
\par
Continuing with the filter definition \eqref{filter-CD-difference-eqn}, suppose we wish to realize a 3-tap filter using two 2-tap filters with complex coefficients. These filters are given by difference equations:
\begin{align} 
y[n] &= p\,x[n] + q\,x[n-1]+ r\,x[n-2], \\
\label{cascading-fil1-eqn} 
    y_1[n] &= a_1\,x[n] + b_1\,x[n-1], \\
\label{cascading-fil2-eqn}
    y_2[n] &= a_2\,y_1[n] + b_2\,y_1[n-1].
\end{align}
On cascading the filters \eqref{cascading-fil1-eqn} and \eqref{cascading-fil2-eqn}, the filter coefficients are related as: $p = a_1a_2$, $q = a_1b_2 + a_2b_1 $, and $r = b_1b_2$. Designing the unitaries for filter-cascaded case involves designing two matrices structured so that their product corresponds to the extended filter realization represented by the resulting matrix. We defined these two matrices as given in Eqs. \eqref{non-u1} and \eqref{non-u2}.
However, $\mathcal{U}_1$ is not a unitary matrix, hence we need to block-encode it. Although $\mathcal{U}_2$ is unitary, in order to compose it with $U_1$, we need to extend $\mathcal{U}_2$ to a larger space. Thus, we introduce the tensor product with Pauli-$Z$-gate as seen in Eq. \eqref{eqn:u-2} to obtain $U_2$. We propose the following design steps.
\\
\textit{Input State}: The signal-encoded qubit state tensored with the additional qubit state (assumed to be initialized to $\ket{0}$) is
    \begin{align}
    % \scalebox{0.9}{$
        \ket{X_n} = \begin{bmatrix}
        1 \\
        0
    \end{bmatrix} \otimes
    \begin{bmatrix}
        x_{n-2} \\
        x_{n-1} \\
        x_n \\
        \sqrt{1-(x_{n-2}^2+x_{n-1}^2+x_{n}^2)}
    \end{bmatrix}.
    % $}
    \end{align} 
\textit{Design of Filter-1 Unitary $U_1$}: Consider the matrix 
    \begin{align} \label{non-u1}
    % \scalebox{0.9}{$
    \mathcal{U}_1 = \begin{bmatrix} 
   \frac{a_1}{\sqrt{a_1^2+b_1^2}} & 0 &  \frac{b_1}{\sqrt{a_1^2+b_1^2}} & 0\\ 
   \frac{b_1}{\sqrt{a_1^2+b_1^2}} & \frac{a_1}{\sqrt{a_1^2+b_1^2}} & 0 & 0\\ 
   0 & \frac{b_1}{\sqrt{a_1^2+b_1^2}} & \frac{a_1}{\sqrt{a_1^2+b_1^2}} &  0 \\
   0 & 0 & 0 & 1
   \end{bmatrix}.
   % $}
    \end{align} 
Next, we block-encode $\mathcal{U}_1$ in $U_1$ given by, \[
    % \scalebox{0.9}{$
    U_1 = \begin{bmatrix}
        \mathcal{U}_1 & \sqrt{I - \mathcal{U}_1 \mathcal{U}_1^\dagger} \phantom{.} \\
        \sqrt{I - \mathcal{U}_1 \mathcal{U}_1^\dagger} & 
        - \mathcal{U}_1
    \end{bmatrix}.
    % $}
    \]
\textit{Design of Filter-2 Unitary $U_2$}: Unitary $\mathcal{U}_2$ realizes the filter-2 computation on two input qubit states, and it is given by,
\begin{align}\label{non-u2}
    % \scalebox{0.9}{$
    \mathcal{U}_2 =  \begin{bmatrix} 
        1 & 0 & 0 & 0 \\ 
        0 & -\frac{a_2}{\sqrt{a_2^2+b_2^2}} & \frac{b_2}{\sqrt{a_2^2+b_2^2}} & 0\\ 
        0 & \frac{b_2}{\sqrt{a_2^2+b_2^2}} & \frac{a_2}{\sqrt{a_2^2+b_2^2}}  & 0 \\ 
        0 & 0 & 0 & 1
    \end{bmatrix}.
    % $}
\end{align}

$U_2$ realizes the overall filter computation at the filter 2 as per \eqref{cascading-fil2-eqn}
and is expressed as: 
\begin{align} \label{eqn:u-2}
    % \scalebox{0.9}{$ 
    U_2 
    = Z \otimes \mathcal{U}_2 
    = \begin{bmatrix}
        1 & 0\\
        0 & -1
    \end{bmatrix} \otimes \mathcal{U}_2
    = \begin{bmatrix}
        \mathcal{U}_2 & 0\\
        0 & - \mathcal{U}_2
    \end{bmatrix}.
    % $}
\end{align}
\textit{Cascading Unitaries $U_2 U_1$}:
Combining the expressions defined above, we apply $U_2 U_1$ to the input state $\ket{X_n}$ such that 
\begin{align}
    U_2 U_1  = \begin{bmatrix} \label{block-u1u2}
        \mathcal{U}_2 \mathcal{U}_1 & \mathcal{U}_2 \sqrt{I - \mathcal{U}_1 \mathcal{U}_1^\dagger} \phantom{.} \\
        -\mathcal{U}_2 \sqrt{I - \mathcal{U}_1 \mathcal{U}_1^\dagger} & \mathcal{U}_2 \mathcal{U}_1
    \end{bmatrix}, 
\end{align}
where $\mathcal{U}_2 \mathcal{U}_1$ has the form,
\begin{align}\label{cascaded-unitaries-eqv}
    \begin{bmatrix}
     * & * & * & 0\\
     * & * & * & 0\\
     \frac{b_1b_2}{\sqrt{(a_1^2+b_1^2)(a_2^2+b_2^2)}} & \frac{a_1b_2+b_1a_2}{\sqrt{(a_1^2+b_1^2)(a_2^2+b_2^2)}} & \frac{a_1a_2}{\sqrt{(a_1^2+b_1^2)(a_2^2+b_2^2)}} & 0 \\
     0 & 0 & 0 & 1
   \end{bmatrix}
\end{align}
From \eqref{cascaded-unitaries-eqv}, the 3rd row of $\mathcal{U}_2 \mathcal{U}_1$ realizes the cascaded filter equivalent result.

Thus, the 3rd or $d$-th row of the upper-left and lower-right blocks of $U$ realize the larger tap filter.
\par 
\textit{Output State $\ket{Y_n}$}: The output qubit state of this cascaded filter system can be expressed as 
\begin{align}\label{cascaded-filter-result}
% \scalebox{0.9}{$
\ket{Y_n} = U_2 U_1 \ket{X_n}.
% $}
\end{align}
\textit{Extending the framework to cascading arbitrary filters quantumly}:
In a general setting where a $d$-tap filter is approximated using $d_1$- and $d_2$-tap filters, $\mathcal{U}_1$ is designed such that for the $d$-th row, entries are zero from column 1 to column $(d_1-1)$, and 
normalized filter coefficients fill up the columns from the $d_1$-th column onwards, followed by zero entries in the remaining columns. For the rows above row $d$, the filter coefficients are left-shifted by one position. Thus a total of $d_1$ rows are fixed in $\mathcal{U}_1$. The design of $\mathcal{U}_2$ remains simple such that only row $d$ is designed like row $d$ of $\mathcal{U}_1$, and the rest of the matrix entries can be designed appropriately. Note that although we have focused on digital signals in this section, analog classical signals can be encoded in hybrid analog-digital quantum systems \cite{liu2024toward}.

\begin{figure}
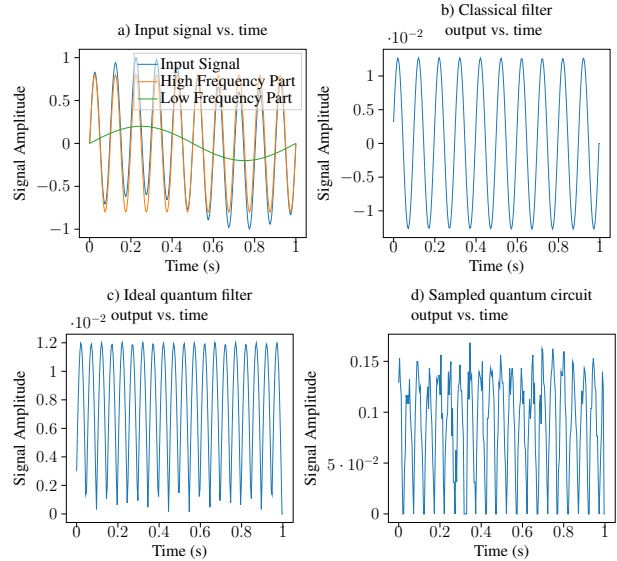

    \scalebox{0.44}{\input{tikz/fig1}\hspace{1mm}\input{tikz/fig2}}
    \vspace{1mm}\\
    \scalebox{0.44}{\input{tikz/fig3}\hspace{1mm}% This file was created with tikzplotlib v0.10.1.
\begin{tikzpicture}

\definecolor{darkgray176}{RGB}{176,176,176}
\definecolor{steelblue31119180}{RGB}{31,119,180}

\begin{axis}[
tick align=outside,
tick pos=left,
title={d) Sampled quantum circuit\\output vs. time}, style={font=\Large},style={align=left},
x grid style={darkgray176},
xlabel={Time (s)},
xmin=-0.05, xmax=1.05,
xtick style={color=black},
y grid style={darkgray176},
ylabel={Signal Amplitude},
ymin=-0.00841432001114766, ymax=0.176700720234101,
ytick style={color=black},
xlabel style={anchor=north},
ylabel style={anchor=south},
label style={font=\Large},
xticklabel style={yshift=5pt}
]
\addplot [semithick, steelblue31119180]
table {%
0 0.128847050800552
0.00401606425702811 0.153093108923949
0.00803212851405622 0.128847050800552
0.0120481927710843 0.116926793336686
0.0160642570281124 0.0441941738241592
0.0200803212851406 0
0.0240963855421687 0
0.0281124497991968 0.0441941738241592
0.0321285140562249 0.0625
0.036144578313253 0.116926793336686
0.0401606425702811 0.108253175473055
0.0441767068273092 0.116926793336686
0.0481927710843373 0.108253175473055
0.0522088353413655 0.132582521472478
0.0562248995983936 0.0988211768802619
0.0602409638554217 0.0826797284707685
0.0642570281124498 0.0698771242968684
0.0682730923694779 0.03125
0.072289156626506 0
0.0763052208835341 0.0541265877365274
0.0803212851405622 0.0988211768802619
0.0843373493975904 0.125
0.0883534136546185 0.149869735103522
0.0923694779116466 0.149869735103522
0.0963855421686747 0.136215591985646
0.100401606425703 0.121030729568982
0.104417670682731 0.125
0.108433734939759 0.0826797284707685
0.112449799196787 0.0698771242968684
0.116465863453815 0.0441941738241592
0.120481927710843 0
0.124497991967871 0
0.1285140562249 0.0541265877365274
0.132530120481928 0.0883883476483184
0.136546184738956 0.11267347735825
0.140562248995984 0.125
0.144578313253012 0.103644524698606
0.14859437751004 0.14320549046737
0.152610441767068 0.132582521472478
0.156626506024096 0.103644524698606
0.160642570281124 0.0883883476483184
0.164658634538153 0.0541265877365274
0.168674698795181 0.03125
0.172690763052209 0
0.176706827309237 0.0541265877365274
0.180722891566265 0.0988211768802619
0.184738955823293 0.132582521472478
0.188755020080321 0.128847050800552
0.192771084337349 0.136215591985646
0.196787148594377 0.136215591985646
0.200803212851406 0.116926793336686
0.204819277108434 0.15625
0.208835341365462 0.09375
0.21285140562249 0.0988211768802619
0.216867469879518 0.03125
0.220883534136546 0
0.224899598393574 0.0541265877365274
0.228915662650602 0.0765465544619743
0.232931726907631 0.09375
0.236947791164659 0.0988211768802619
0.240963855421687 0.139754248593737
0.244979919678715 0.116926793336686
0.248995983935743 0.136215591985646
0.253012048192771 0.149869735103522
0.257028112449799 0.0765465544619743
0.261044176706827 0.103644524698606
0.265060240963855 0.03125
0.269076305220883 0.03125
0.273092369477912 0
0.27710843373494 0.0625
0.281124497991968 0.03125
0.285140562248996 0.103644524698606
0.289156626506024 0.125
0.293172690763052 0.128847050800552
0.29718875502008 0.146575492494482
0.301204819277108 0.121030729568982
0.305220883534137 0.121030729568982
0.309236947791165 0.09375
0.313253012048193 0.0765465544619743
0.317269076305221 0
0.321285140562249 0
0.325301204819277 0
0.329317269076305 0
0.333333333333333 0.103644524698606
0.337349397590361 0.139754248593737
0.34136546184739 0.11267347735825
0.345381526104418 0.168286400222953
0.349397590361446 0.132582521472478
0.353413654618474 0.136215591985646
0.357429718875502 0.108253175473055
0.36144578313253 0.0698771242968684
0.365461847389558 0.03125
0.369477911646586 0.03125
0.373493975903614 0
0.377510040160643 0.0541265877365274
0.381526104417671 0.09375
0.385542168674699 0.125
0.389558232931727 0.139754248593737
0.393574297188755 0.136215591985646
0.397590361445783 0.132582521472478
0.401606425702811 0.121030729568982
0.405622489959839 0.149869735103522
0.409638554216867 0.0765465544619743
0.413654618473896 0.0698771242968684
0.417670682730924 0.03125
0.421686746987952 0
0.42570281124498 0
0.429718875502008 0.09375
0.433734939759036 0.09375
0.437751004016064 0.108253175473055
0.441767068273092 0.136215591985646
0.44578313253012 0.14320549046737
0.449799196787149 0.136215591985646
0.453815261044177 0.139754248593737
0.457831325301205 0.125
0.461847389558233 0.11267347735825
0.465863453815261 0.0541265877365274
0.469879518072289 0
0.473895582329317 0
0.477911646586345 0.0625
0.481927710843373 0.0826797284707685
0.485943775100402 0.128847050800552
0.48995983935743 0.125
0.493975903614458 0.15625
0.497991967871486 0.14320549046737
0.502008032128514 0.128847050800552
0.506024096385542 0.128847050800552
0.51004016064257 0.0988211768802619
0.514056224899598 0.0625
0.518072289156627 0.0441941738241592
0.522088353413655 0
0.526104417670683 0.03125
0.530120481927711 0.0826797284707685
0.534136546184739 0.103644524698606
0.538152610441767 0.14320549046737
0.542168674698795 0.116926793336686
0.546184738955823 0.136215591985646
0.550200803212851 0.14320549046737
0.554216867469879 0.132582521472478
0.558232931726908 0.0988211768802619
0.562248995983936 0.0826797284707685
0.566265060240964 0.03125
0.570281124497992 0
0.57429718875502 0
0.578313253012048 0.0625
0.582329317269076 0.0826797284707685
0.586345381526104 0.103644524698606
0.590361445783132 0.11267347735825
0.594377510040161 0.14320549046737
0.598393574297189 0.128847050800552
0.602409638554217 0.108253175473055
0.606425702811245 0.121030729568982
0.610441767068273 0.0698771242968684
0.614457831325301 0.0625
0.618473895582329 0.03125
0.622489959839357 0
0.626506024096385 0.0625
0.630522088353414 0.0883883476483184
0.634538152610442 0.11267347735825
0.63855421686747 0.09375
0.642570281124498 0.108253175473055
0.646586345381526 0.146575492494482
0.650602409638554 0.15625
0.654618473895582 0.108253175473055
0.65863453815261 0.128847050800552
0.662650602409638 0.0625
0.666666666666667 0.0441941738241592
0.670682730923695 0
0.674698795180723 0
0.678714859437751 0.03125
0.682730923694779 0.108253175473055
0.686746987951807 0.116926793336686
0.690763052208835 0.0883883476483184
0.694779116465863 0.162379763209582
0.698795180722891 0.159344359799775
0.70281124497992 0.128847050800552
0.706827309236948 0.125
0.710843373493976 0.103644524698606
0.714859437751004 0.0765465544619743
0.718875502008032 0.03125
0.72289156626506 0
0.726907630522088 0.0541265877365274
0.730923694779116 0.0826797284707685
0.734939759036145 0.116926793336686
0.738955823293173 0.136215591985646
0.742971887550201 0.14320549046737
0.746987951807229 0.162379763209582
0.751004016064257 0.153093108923949
0.755020080321285 0.132582521472478
0.759036144578313 0.121030729568982
0.763052208835341 0.0826797284707685
0.767068273092369 0
0.771084337349398 0
0.775100401606426 0.0441941738241592
0.779116465863454 0.0541265877365274
0.783132530120482 0.108253175473055
0.78714859437751 0.139754248593737
0.791164658634538 0.136215591985646
0.795180722891566 0.136215591985646
0.799196787148594 0.159344359799775
0.803212851405622 0.159344359799775
0.807228915662651 0.108253175473055
0.811244979919679 0.09375
0.815261044176707 0.03125
0.819277108433735 0
0.823293172690763 0
0.827309236947791 0.0441941738241592
0.831325301204819 0.0765465544619743
0.835341365461847 0.0826797284707685
0.839357429718875 0.103644524698606
0.843373493975904 0.149869735103522
0.847389558232932 0.146575492494482
0.85140562248996 0.128847050800552
0.855421686746988 0.108253175473055
0.859437751004016 0.0883883476483184
0.863453815261044 0.0765465544619743
0.867469879518072 0.03125
0.8714859437751 0
0.875502008032128 0.03125
0.879518072289157 0.0698771242968684
0.883534136546185 0.09375
0.887550200803213 0.121030729568982
0.891566265060241 0.108253175473055
0.895582329317269 0.125
0.899598393574297 0.14320549046737
0.903614457831325 0.139754248593737
0.907630522088353 0.125
0.911646586345381 0.0883883476483184
0.91566265060241 0.03125
0.919678714859438 0
0.923694779116466 0
0.927710843373494 0.0698771242968684
0.931726907630522 0.103644524698606
0.93574297188755 0.103644524698606
0.939759036144578 0.14320549046737
0.943775100401606 0.153093108923949
0.947791164658634 0.14320549046737
0.951807228915663 0.125
0.955823293172691 0.121030729568982
0.959839357429719 0.0765465544619743
0.963855421686747 0.0765465544619743
0.967871485943775 0.03125
0.971887550200803 0
0.975903614457831 0.03125
0.979919678714859 0.0625
0.983935742971887 0.0826797284707685
0.987951807228916 0.139754248593737
0.991967871485944 0.125
0.995983935742972 0
1 0
};
\end{axis}
\end{tikzpicture}}
    \vspace*{-0.5\baselineskip}
    \caption{Simulation results for a symmetric 3-tap high pass filter with low and high-frequency signal at the input. b)-d) depict the output of the classical filter, ideal quantum filter and quantum machine-simulator quantum filter.}
    \vspace*{-0.5\baselineskip}
    \label{fig:simulation-results}
    \vspace*{-1\baselineskip}
\end{figure}

\textit{Simulation Results:} To validate our construction, Fig. \ref{fig:simulation-results} compares classical and quantum filter outputs. We apply a 3-tap high-pass filter, $y[n] = -\frac{1}{4} x[n]+\frac{1}{2} x[n-1]-\frac{1}{4} x[n-2]$, to an input signal consisting of both high and low-frequency components. The classical filter output, ideal quantum output and measured quantum machine output are calculated for the given signal. The output Fig. \ref{fig:simulation-results}d was obtained on IBM Qiskit's QASM simulator for 1024 shots \cite{QiskitWebSite}. The reconstruction errors in Fig. \ref{fig:simulation-results}d result from statistical fluctuations inherent to the probabilistic nature of
quantum computers due to the finite number of measurements. It can be observed that the measurement outcomes in panel \ref{fig:simulation-results}d match those of panel \ref{fig:simulation-results}c. We note that while any complex signal amplitude and phase can be encoded as a quantum state,
when measured under the $Z$-basis, only the modulus of the signal can be extracted, and hence rectified in nature. To obtain complete information on the phase of the amplitude, measurement under other bases such as the $X$-basis is required as well. We leave the efficient design of measurement operators for obtaining both phase and amplitude for future work.
%%%%%****************************%%%%%%%
%%%%%%%%%%%%%%%%%%%%%%%%%%%%%%%%%%%%

\section{Quantum to Classical}
\label{sec:quantum2classical}

Current NISQ quantum computers are more susceptible to quantum noise compared to their classical counterparts, which is a major obstacle to implementing large and complicated quantum circuits \cite{Preskill2018quantumcomputingin}. 
As generic quantum states cannot be fanned out like classical bits, there is a need to perform signal processing of quantum states on classical systems.
\par
%Digital Quantum States to Digital Signals%
The versatility of qubits lies in their ability of superposition and entanglement, which can be harnessed to process discrete or continuous classical data simultaneously. However, measuring the qubit results in collapsing the wave function to a definite outcome, and multiple measurements are required to estimate the amplitudes. Drawing insights from the statistical nature of the measurements can lead to quantum-statistical signal processing based on the %existing
literature on classical statistical signal processing \cite{gray2004introduction,10.5555/151045}.
\par
A major part of current quantum frameworks consists of quantum error correction schemes \cite{RevModPhys.95.045005} that aim at correcting errors and mitigating the effects of quantum noise. Given that qubits are inherently noisy, converting quantum data to classical can also suffer from state preparation and measurement error \cite{Geller_2021,Harper_2020}. Therefore, performing classical processing of noisy samples drawn from quantum systems would further require tailored classical error correction and mitigation schemes.

% Analog Quantum States to Analog Signals?
\par 
Apart from digital signals, analog quantum resources are indispensable \cite{liu2024toward}. Techniques such as homodyne detection \cite{15986} can convert quantum analog signals to classical analog ones. However, entanglement in analog quantum systems may result in classical analog signals that are challenging to process using existing classical signal processing techniques.  It remains to understand the limitations on the range and boundaries of analog properties and filter design techniques that can be realized on either classical or quantum systems.  

%%%%%%%%%%%%%%%%%%%%%%%%%%%%%%%%%%%%
\section{Discussion and Conclusions}
\label{sec:dis-conclusion}
In this work, we present a framework for performing classical signal processing on quantum systems by developing an array of signal encoding and unitary design techniques. We realize classical filter design and filter cascading on quantum systems by cascading block-encoded lower-order filters, designed as unitary operators on quantum computers. We discuss how to perform classical to quantum conversion and vice-versa for applications where performing classical signal processing on quantum systems can lead to speed-ups. We also discuss the motivations and challenges of converting quantum signals to the classical domain. Quantifying the circuit complexity of our constructions is currently ongoing. Overall, by connecting classical and quantum signal processing, we hope this work serves as a foundation for future development and cross-fertilization of the two fields.

%%%%%%%%%%%%%%%%%%%%%%%%%%%%%%%%%%%%

%\section*{Acknowledgments}
%BNB was supported in part by a Simons Foundation grant No. 584741.

\newpage

\balance
\bibliographystyle{IEEEtran}
\bibliography{IEEEabrv,ref}

\end{document}